# Entropy production and Lyapunov instability at the onset of turbulent convection

V. M. Castillo and Wm. G. Hoover

*Department of Applied Science, University of California at Davis–Livermore and Lawrence Livermore National Laboratory,
Livermore, California 94551-7808*

Computer simulations of a compressible fluid, convecting heat in two dimensions, suggest that, within a range of Rayleigh numbers, two distinctly different, but stable, time-dependent flow morphologies are possible. The simpler of the flows has two characteristic frequencies: the rotation frequency of the convecting rolls, and the vertical oscillation frequency of the rolls. Observables, such as the heat flux, have a simple-periodic (harmonic) time dependence. The more complex flow has at least one additional characteristic frequency—the horizontal frequency of the cold, downward- and the warm, upward-flowing plumes. Observables of this latter flow have a broadband frequency distribution. The two flow morphologies, at the same Rayleigh number, have different rates of entropy production and different Lyapunov exponents. The simpler "harmonic" flow transports more heat (produces entropy at a greater rate), whereas the more complex "chaotic" flow has a larger maximum Lyapunov exponent (corresponding to a larger rate of phase-space information loss). A linear combination of these two rates is invariant for the two flow morphologies over the entire range of Rayleigh numbers for which the flows coexist, suggesting a relation between the two rates near the onset of convective turbulence. [S1063-651X(98)04612-1]

## I. INTRODUCTION

In order to transport heat more effectively, a fluid spontaneously makes a transition from quiescent, Fourier heat conduction to convection—where currents convey warm fluid to the cold boundary and cool fluid to the hot boundary—at a sufficiently high Rayleigh number (dimensionless temperature gradient). At a much higher Rayleigh number, the system makes another transition, from steady to time-dependent convection. This transition was noted by Clever and Busse [1] for an incompressible fluid, and by Rapaport [2] using molecular dynamics. The steady-unsteady transition occurs at Ra≈8×10⁴ for the fluid discussed here, and marks the beginning of a periodic motion in which the rolls oscillate vertically. Eventually, this periodic motion gives way to chaotic flow as the vertical thermal plumes start to sweep from side to side.

Our simulations of the fully compressible Navier-Stokes equations for a viscous, heat-conducting fluid enclosed between two rigid, thermal boundaries in the presence of a body force reveal that, within a range of Rayleigh numbers, both laminarlike harmonic flow and turbulentlike chaotic flow morphologies are stable solutions for the same Rayleigh number. Figure 1 shows time sequences for the laminarlike (left hand side) and chaotic (right hand side) flows—both at the same Rayleigh number $2\times10^5$. The sequence of the harmonic flow shows the thermal plumes oscillating vertically. This frequency is in addition to the frequency of the fluid moving around the two counter-rotating convection cells. The harmonic flow is more effective in transporting heat. The chaotic flow has at least one additional characteristic frequency—that of the plumes sweeping from side to side, disturbing the thermal boundary layer on the opposite side. The horizontal sweeping near the opposite thermal boundary causes flow in a direction counter to the net flow of heat, resulting in a less efficient heat transfer. The addition of a third incommensurate frequency is, according to New-house, Ruelle, and Takens's idea of the route to chaos [3], enough to induce highly unstable chaotic motion.

The timed-averaged Nusselt number (dimensionless heat flux) is plotted in Fig. 2 for flows with $8\times10^4 \lesssim \mathrm{Ra} \lesssim 5\times10^5$. Within the dual-morphology region, the Nusselt number for the two-frequency periodic flow (higher) and the three-frequency chaotic flow (lower) are joined by a hyster-esis loop. The upper branch of the loop continues from the steady-state region into the periodic region. The lower branch continues from the chaotic region to fully developed turbulent convection. As the Rayleigh number is increased quasistatically from a steady state, periodic convection develops first. As the Rayleigh number is increased further, the system makes a transition to chaotic flow. The Nusselt number, as seen in Fig. 2, shows an initial drop along this path, followed by an increase with increasing Ra. The term "quasistatically" is used to mean that a well-developed state at a given Rayleigh number is used as an initial condition for a run at a slightly different Rayleigh number. The simulation is allowed to run for several thousand sound traversal times to eliminate transient effects. The Rayleigh number can be varied in our simulations by changing either the transport coefficients (which also varies the diffusion traversal times but not the sound traversal time), the temperature of the hot boundary (which also varies the sound traversal time and the thermal expansion coefficient), or the length scale (which varies the diffusion and sound traversal times). In all three cases, a hysteresis is present, but when the length scale is varied, the range of the dual-morphology region is decreased.

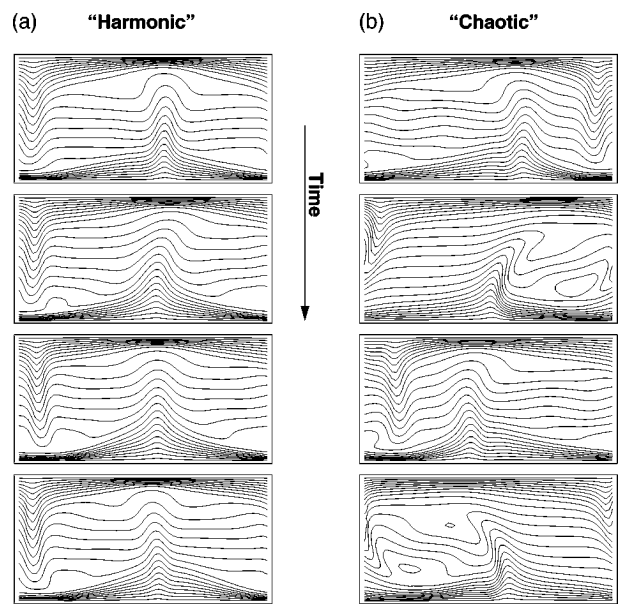

FIG. 1. Temperature contours for harmonic and chaotic flows at Ra=200 000. The time sequence appearing on the left shows the flow for "harmonic" convection. Observables, such as the Nusselt number, vary harmonically in time as the plumes penetrate the opposite boundary layers. There are two characteristic frequencies in this flow—that of the fluid being carried by the convection rolls and that of the vertical oscillation of the plumes. The chaotic flow (sequence appearing on the right) has at least one additional frequency—that of the plumes sweeping side to side.

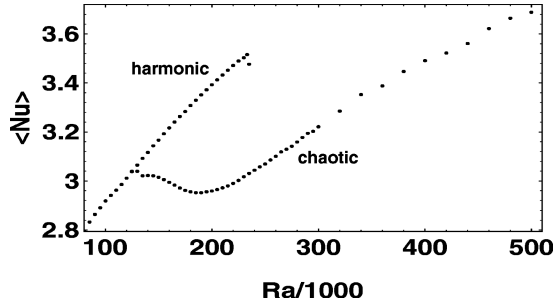

FIG. 2. Time-averaged Nusselt and Rayleigh number for flows in the dual-morphology region. The harmonic flows transport more heat.

Corresponding experiments of turbulent convection using mercury, a low-Prandtl-number (dimensionless ratio of viscosity to heat-conductivity) fluid, reported in Ref. [4], shows a "bump" in the Nusselt number–Rayleigh number relation where the Nusselt number drops, at a point, with increasing Rayleigh number. This drop in Nusselt number is accompanied by an apparent change in flow morphology, indicated by the temperature-fluctuation histogram at a probe fixed at the center of the cell. The temperature histogram goes from a dual-maximum profile before the Nusselt number drop to a single-maximum profile and has been interpreted, in the absence of the ability to visualize the actual flow, as a change in the number of convection rolls in the system. This change in the temperature history profile is also seen in in our simulations, and corresponds to the transition from harmonic to turbulentlike flow. Another set of experiments, involving the convection of gaseous helium [5], describes transitions in the flow morphology for Rayleigh numbers from $10^3$ to $10^{11}$. The onset of the "oscillatory" convecting flow is reported for $Ra = 9 \times 10^4$. Chaotic flow is reported for $1.5 \times 10^5 \leq Ra \leq 2.5 \times 10^5$. A drop in the Nusselt number is seen in the data for the transition from the oscillatory to chaotic flow. Even though these experiments were not conducted to investigate the existence of a hysteresis in the Nusselt number or the presence of dual morphologies, a change in morphology is noted in both, and the drop in Nu can be seen in the plots of the data, as discussed in Ref. [6]. The drop in the Nusselt number at the transition from periodic to chaotic convection, present in distinctly different systems—two dimensional (2D) and 3D, low and high Prandtl numbers, and compressible and nearly incompressible fluids—are apparently universal, and support the existence of a hysteretic loop joining the two morphologies near this transition. Dual morphology has been reported in steady convection flows for both compressible [7] and incompressible [1] fluids.

## II. ENTROPY PRODUCTION AND THE LOSS OF PHASE-SPACE INFORMATION

Because the temporally periodic flow transports more heat, it must likewise produce entropy at a greater rate. The total internal entropy production is the integral over the system volume of the internal entropy production per unit volume $g_i$. According to local thermodynamic equilibrium with linear transport theory, $g_i$ is defined

$$\dot{S}_i = \int_V g_i \, dV, \quad g_i \equiv -\frac{\vec{q} \cdot \nabla T}{T^2} + \frac{\overline{\overline{\sigma}} : \nabla \vec{u}}{T},$$

where $\vec{q}$ is the heat flux and $\overline{\overline{\sigma}}$ is the stress tensor. This total internal entropy production can be made dimensionless by dividing it by the internal entropy production for the equivalent quiescent system (with heat conduction only),

$$\dot{S}_{ND} \equiv \dot{S}_i / \dot{S}_F, \quad \dot{S}_F = -W \int_0^L dy \frac{\vec{q}_F \cdot \nabla T_F}{T^2},$$

where $\vec{q}_F = -\kappa \nabla T_F$, $\nabla T_F = \vec{y} \Delta T / L$, and $W$ is the width of the system. Figure 3(a) shows the time-averaged dimensionless internal entropy production for various Rayleigh number flows including the dual-morphology region. The dimensionless entropy production is equivalent to the Nusselt number (time-averaged quantities) in these simulations in the $\Delta x \rightarrow 0$ limit.

The chaotic flow loses phase-space information at a greater rate. The Kolmogorov entropy, the sum of the positive Lyapunov exponents $\Sigma \lambda^+$, quantifies this rate. For our simulations, however, it is not easy to obtain the entire Lyapunov spectrum, and it is also not clear how well the spectrum for our discrete approximation describes the continuous system. Instead, we calculate the maximum Lyapunov exponent and use it as an estimate of the rate of loss of phase-space information. To make this rate dimensionless, the maximum Lyapunov exponent $\lambda_1$ is multiplied by the sound traversal time $\tau_s$, the time for information to traverse the height of the system. Figure 3(b) shows the dimensionless rate at which phase-space information is lost, $\lambda_1 \tau_s$, for various systems. In this case, the hysteretic loop is also evident, but with the upper branch now representing the rate for chaotic and turbulent flows.

From Figs. 3(a) and 3(b), one can see that the difference in the dimensionless entropy production rates corresponds to the difference in the dimensionless rate of loss of phase-space information for flows at the same Rayleigh number. This implies a connection between the thermodynamic en-tropy production and the information entropy rate near the onset of convective turbulence. A linear sum of the two rates [see Fig. 3(c)] connects the two morphologies over the entire range for which they coexist. With the appropriate choice of the linear coefficient, the hysteresis in the linear sum is greatly reduced so that the linear sum appears to increase continuously with Ra.

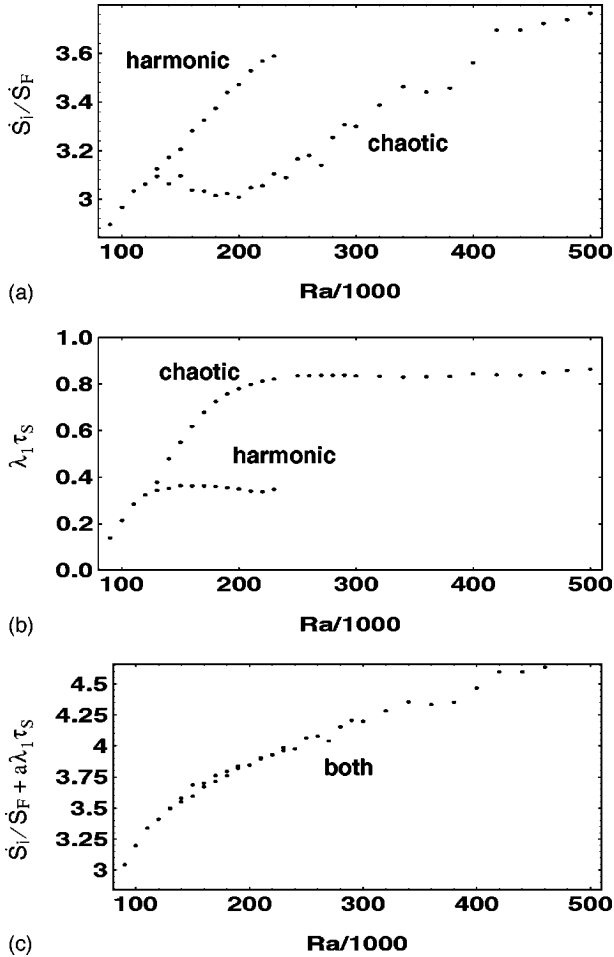

FIG. 3. Entropy production rates for harmonic and chaotic flows. (a) The time-averaged dimensionless internal entropy production rate. The harmonic flow produces entropy at a greater rate. (b) The dimensionless maximum Lyapunov exponent. The chaotic flow loses phase-space information at a greater rate. (c) A linear sum of the two dimensionless rates connects the harmonic and chaotic flows over the entire range for which they coexist.

For higher Rayleigh numbers, within the turbulent region ($2.5 \times 10^5 <$ Ra), a scaling relation between this linear combination and the Rayleigh number, $\dot{S}_{ND} + a\lambda_1 \tau_s \sim$ Ra$^{2/9}$, is suggested by the data. This relation is somewhat inconclusive because the data for our simulations only range a little over one decade. Figure 4 shows a log-log plot of this linear combination as a function of the Rayleigh number.

A scaling relation between the dimensionless heat flux (equal to the dimensionless entropy production) and the Rayleigh number has been the subject of much investigation. Experimental studies of convecting helium gas [8] demonstrate the existence of a scaling region where the convective part of the heat flux is related to the Rayleigh number by Nu$-1 \sim$ Ra$^{0.282 \pm 0.006}$. This is different from the "classical" result [9] relating the heat flux to Ra$^{1/3}$, which is based on the assumption of marginal stability at a thermal boundary layer. Numerical simulations of incompressible Boussinesq fluids [10,11] also reveal a scaling relation close to the experimental one for Rayleigh numbers between $10^8$ and $10^{15}$.

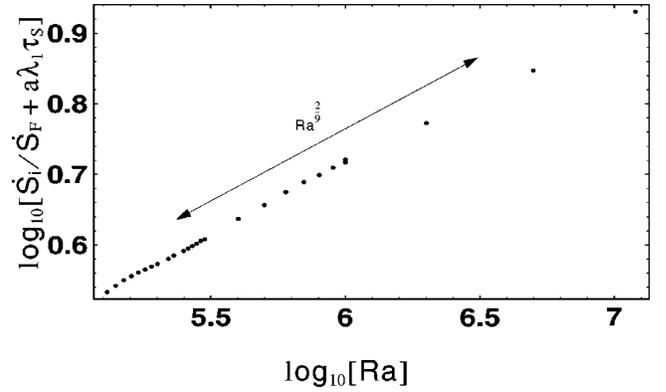

FIG. 4. The dimensionless rate sum vs the Rayleigh number on a doubly logarithmic scale.

The "classical" result relating the dimensionless heat flux to Ra$^{1/3}$ is based on the assumption of a thin thermal boundary layer across which the time-averaged temperature changes by $\Delta T/2$, where $\Delta T$ is the change in temperature across the system. The thickness of the boundary layers, $\delta$, is such as to just deny the instability leading to local convection. The "local" critical Rayleigh number Ra$_c$ $= \rho g \delta^3/\eta\kappa(\Delta T/2)$, and the Nusselt number Nu, can be computed as the Fourier conduction across the boundary layer, $q_{bl} = -\kappa(\Delta T/2)/\delta$ divided by the conduction across the system, $q_F = -\kappa(\Delta T)/L$, where $L$ is the height of the system. By solving for $\delta$ in terms of Nu, one obtains Nu $\sim$ (Ra$/$Ra$_c)^{1/3}$. This power-law relation has been developed independently in at least two different ways [12,13].

There are a number of theories that account for the deviation from the classic $\frac{1}{3}$ value in the hard turbulence region. Castaing et al. [8] proposed that a large scale fluctuating wind may stabilize the thermal boundary layer to yield the $\frac{2}{7}$ power-law relation. She [14] analyzed the Boussinesq equations with the assumption that the central fluctuating temperature field interacts strongly with the turbulent velocity field. Shraiman and Siggia [15] analyzed the Boussinesq equations with the assumption that thermal boundary layer is nested within the viscous boundary layer. Ching [16] used the assumption that a large nonuniform shear influences the thickness of the thermal boundary layer.

The existence of this $\frac{2}{7}$ power-law relation between the Rayleigh number and the dimensionless heat flux, along with an exponential temperature distribution, are the main indicators of "hard" turbulence. This has been reported for two-dimensional convection of a Boussinesq fluid [10,11,17], and is evident in our simulations for chaotic flows [6].

## III. COMPUTATIONAL DETAILS

To study the transition to turbulent convection, we simulate the fully compressible Navier-Stokes equations for a fluid with an ideal gas equation of state, $P_{eq} = \rho k_B T = \rho e$, enclosed between two rigid thermal boundaries separated by a distance $L$, and in the presence of a body force $g$. The vertical boundaries are periodic and have a length scale corresponding to a cell with an aspect ratio of 2. With the Boltzmann constant, the mean density, and the isochoric heat capacity set to unity, the Rayleigh number for the system is defined as Ra$= \alpha g \Delta T L^3 / \eta \kappa$ where $\alpha$ is the thermal expansion coefficient, and $\eta$ and $\kappa$ are shear viscosity and heat transfer coefficients. Because, for an ideal gas, $\alpha = T^{-1}$, and the body force can be assigned a magnitude such that a small volume element of fluid moving from the lower, high-temperature boundary to the upper, low-temperature boundary gains a potential energy to exactly compensate for the loss in thermal energy ($g = \Delta T / L$), the Rayleigh number can be written as

$$\text{Ra} = \frac{\Delta T^2 L^2}{T \eta \kappa}.$$

The Prandtl number Pr is set to unity for all the simulations.

### A. Numerical methods

The continuum equations for the time development of the density, velocity, and internal energy per unit mass $\{\rho, \vec{u}, e\}$ can be written in terms of the Eulerian derivatives at fixed space locations:

$$\partial \rho / \partial t = -\nabla \cdot (\rho \mathbf{u}),$$

$$\partial \mathbf{u} / \partial t = -\mathbf{u} \cdot \nabla \mathbf{u} + (1/\rho) \nabla \cdot \bar{\bar{\sigma}} + \mathbf{g},$$

$$\partial e / \partial t = -\mathbf{u} \cdot \nabla e + (1/\rho) [\nabla \mathbf{u} : \bar{\bar{\sigma}} - \nabla \cdot \mathbf{q}].$$

It is, however, desirable to express the finite-difference form of these equations in conservative form with the mass density, momentum, and total energy calculated from the corresponding fluxes. Otherwise, the simulations can become unstable at long times. To avoid the shortest-wavelength even-odd instability, a ''dual'' grid is used where the momentum and total energy are updated on one grid and the density is updated on the other. Along horizontal or vertical directions, the fluxes of these conserved quantities are calculated on the other grid.

Piecewise cubic interpolating polynomials, ''cubic splines,'' are used to determine the value of the state variables and their gradients on the other grid with high accuracy (error of order $\Delta x^4$ in the bulk and $\Delta x^2$ near the rigid boundaries). For simulations with a grid spacing smaller than the thermal and viscous boundary layer length, the lower-order error is not believed to effect the rest of the calculations because the thermal and viscous transport are linear in that region. Because the state variables are known at the vertices of an equally spaced grid, the midpoint interpolant and gradient reduce to

$$\hat{f}_0 = \frac{f_+ + f_-}{2} - \frac{h^2}{16}(M_+ + M_-) + O(h^4),$$

$$f_0{}' = \frac{f_+ - f_-}{h} - \frac{h}{24}(M_+ - M_-) + O(h^4),$$

where $M_i$ is the second derivative of the spline function at $x_i$. The classic fourth-order Runga-Kutta method is used to integrate the equations in time. The time step in the simulations is governed by the Courant condition, taking into ac-count the sound speed, so that sound waves may be resolved. For these simulations, periodic boundaries are used on the sides and rigid, no-slip, constant temperature conditions are used for the upper and lower boundaries.

The viscous stress tensor $\bar{\bar{\sigma}}$, in $d$ dimensions, is defined

$$\bar{\bar{\sigma}} = \left( \eta_v - \frac{2}{d} \eta \right)(\nabla \cdot \mathbf{u}) \bar{\bar{I}} + \eta (\nabla \mathbf{u} + \nabla \mathbf{u}^t) - P_{\text{eq}} \bar{\bar{I}},$$

where the equilibrium pressure is defined by the ideal gas equation of state $P_{\text{eq}} = \rho e$, $\eta$ is the shear viscosity, $\eta_v$ is the bulk viscosity, and $d$ is the dimension of the system. The system can be further simplified by allowing the bulk viscosity to vanish, but the dimension of the system must be carefully considered. Although these simulations are computed for a two-dimensional system ($d = 2$), we desire that quantities such as the viscous work $\nabla \mathbf{u} : \bar{\bar{\sigma}}$ be equivalent to those of a three-dimensional system with a vanishing bulk viscosity. Therefore, we set $\eta_v = \eta / 3$.

### B. Internal entropy production

The total internal entropy production is calculated by starting with a well-established convection state and averaging the volume integral for over $10^3$ sound traversal times. In Fig. 3(a), the upper branch of the hysteresis loop represents the simple, periodically convecting systems, and is generated by starting with steady convection and increasing the Rayleigh number quasistatically. This simple periodic mode is about 10% more efficient at transporting heat. This efficient flow continues to a critical Rayleigh number where the Nusselt number drops and the flow becomes turbulent. The turbulent flow morphology continues as the Rayleigh number is increased. It should be noted here that in this turbulent regime, the dimensionless entropy production (identically equal to the Nusselt number) is related to the Rayleigh number by the $\frac{2}{7}$ power law $\dot{S}_{ND} \sim \text{Ra}^{2/7}$. As the Rayleigh number is decreased from this point, the lower branch of this plot is generated as the flow remains somewhat turbulent. It is possible to generate either of these two different flow morphologies by starting with different initial conditions within this hysteresis range. Toward the center of the hysteresis range, Ra$\approx 2 \times 10^5$, the different flows are insensitive to rather large perturbations (on the order of 10%). Near the transitional Rayleigh number, the efficient flow is sensitive to perturbation. This hysteresis persists as the grid is refined, and is therefore expected to exist in the continuum limit.

### C. Lyapunov exponent

Our discrete approximation of the system uses a finite number of variables to approximate its state. Each state variable can be thought of as a degree of freedom in a multidimensional dynamical system whose set of Lyapunov exponents $\{\lambda_i\}$ has a cardinality equal to the number of state variables. Examples of Lyapunov spectra for various systems can be seen in Ref. [18]. The spectrum describes the rates at which the phase-space volume grows or decays in time. The maximum Lyapunov exponent $\lambda_1$ indicates the chaotic nature of the system by indicating whether neighboring points in phase space converge (for a stable system), diverge (for a chaotic system), or remain close neighbors (for a neutrally stable system) in general:

$$\lambda_1 \equiv \lim_{t \to \infty} \frac{1}{t} \ln \frac{|\Delta(t)|}{|\Delta(0)|},$$

$$|\Delta(0)| \to 0.$$

The maximum Lyapunov exponent is also relatively easy to compute. From a well-established state the maximum Lyapunov exponent is calculated. This is done by selecting a state (at random) that represents a small perturbation, and allowing both to evolve for a time step before uniformly adjusting all of the state variables so that the phase-space distance is equal to the initial perturbation [19]. By doing this several thousand times, the perturbation direction rotates to give the eigenvector corresponding to the maximum Lyapunov exponent. Once this alignment is established, the time-averaged Lagrange multiplier needed to readjust the phase-space distance is exactly equal to $\lambda_1$. Figure 3(b) shows $\lambda_1 \tau_s$ for various Rayleigh numbers. Five hundred sound traversal times are allowed to pass for the perturbation to establish itself, and another 500 sound traversal times are used for the time averaging. The hysteresis loop, in this plot, goes counterclockwise. As the Rayleigh number is increased, from a steady state, the lower branch is generated until the transition to turbulence, which is indicated by a change in the flow morphology and a sudden jump in the dimensionless maximum Lyapunov exponent. As the Rayleigh number is decreased, $\lambda_1 \tau_s$ smoothly decreases along the upper branch of the hysteresis loop.

## IV. CONCLUSIONS

In this paper, we report on time-dependent convective flows, with Rayleigh numbers in a region near the transition to chaotic convection, in which two distinctly different, but stable, flow morphologies coexist. A relationship is revealed between two important quantities that characterize these non-equilibrium, time-dependent flows—the entropy production rate and the rate at which phase-space information is lost. These relations are based on well-established flow simulations, so that the transient effects are minimized. The flows are also, in general, stable and insensitive to random perturbations.

The existence of this dual-morphology region and the hysteresis loop in the dimensionless heat flux that connects the two flow morphologies is also corroborated by the existence of sharp discontinuities in Nusselt number data reported for two independent experiments with two very different fluids—gaseous helium and liquid mercury. These experiments not only support these claims, but suggest a universality for this phenomena.

The scaling relation between the Rayleigh number and the linear sum of the entropy production rate and the $K$ entropy (rate) may shed light on the continuing investigation on the relation between the dimensionless heat flux and the Rayleigh number. Recently, a power-law relation between the Lyapunov exponent times a characteristic time of the system (like a sound traversal time) and the Reynolds number was proposed [20]: $\langle \lambda \rangle \tau_0 \sim \mathrm{Re}^{0.459}$. This, along with the relation between the Reynolds number and the Rayleigh number, further supports the work detailed in this paper.

Currently, a model of a double pendulum, with thermally conducting and thermally expanding masses immersed in a constant-gradient temperature field, is being studied [21]. The pendulum assists in transporting heat in a way that is analogous to buoyancy-driven convection. Both a Rayleigh-like order parameter and a dimensionless entropy production rate can be defined for this system. Initial results suggest a drop in the entropy production rate at the transition to chaos, although a dual-mode region is not evident.

## ACKNOWLEDGMENT

This work was performed under the auspices of the U.S. Department of Energy through University of California Contract No. W-7405-Eng-48.